\documentclass[aps,pre,preprint,groupedaddress]{revtex4-1}

\usepackage{graphicx}

\begin{document}


\title{Field theoretical prediction of a property of the tropical cyclone}


\author{F. Spineanu}
\author{M. Vlad}
\affiliation{National Institute of Laser, Plasma and Radiation Physics\\
Magurele, Bucharest 077125, Romania}


\date{\today}

\begin{abstract}
The large scale atmospheric vortices (tropical cyclones, tornadoes) are
complex physical systems combining thermodynamics and fluid-mechanical
processes. The well known tendency of vorticity to self-organization, an
universal property of the two-dimensional fluids, is part of the full
dynamics, but its description requires particular methods. The general
framework for the thermodynamical and mechanical processes is based on
conservation laws while the vorticity self-organization needs a variational
approach. It is difficult to estimate to what extent the vorticity
self-organization (a purely kinematic process) have influenced the
characteristics of the tropical cyclone at stationarity. If this influence
is substantial it is expected that the stationary state of the tropical
cyclone has the same nature as the vortices of many other systems in nature:
ideal (Euler) fluids, superconductors, Bose - Einstein condensate, cosmic
strings, etc. 

In previous works we have formulated a description of the $2D$ vorticity self-organization in
terms of a classical field theory. It is compatible with the more
conventional treatment based on conservation laws, but the field theoretical
model reveals properties that are almost inaccessible to the conventional
formulation: it identifies the stationary states as being close to
self-duality. This is of highest importance: the self-duality is at
the origin of all coherent structures known in natural systems. Therefore
the field theoretical (FT) formulation finds that the cuasi-coherent form of
the atmospheric vortex (tropical cyclone) at stationarity is an expression
of this particular property. Since the FT model is however limited to the
self-organization of the vorticity and does not cover the full dynamics, one
still needs to quantify the relative importance of these processes.

In the present work we examine a strong property of the tropical cyclone,
which arises in the FT formulation in a natural way: the equality of the
masses of the particles associated to the matter field and respectively to
the gauge field in the FT model is translated into the equality between the
maximum radial extension of the tropical cyclone and the Rossby radius. For
the cases where the FT model is a good approximation we calculate
characteristic quantities of the tropical cyclone and find good comparison
with observational data.
\end{abstract}


\maketitle



\section{Introduction}

The objective of the present work is the derivation of a property of the
large scale stationary vortical structures in atmosphere, in particular the
stationary state of the tropical cyclone: the maximal radial extension of
the vortex is equal to the characteristic spatial dimension in the problem,
the Rossby radius. The derivation is formulated within a model of the
self-organization of vorticity in $2D$ flows and requires an introductory
explanation.

\bigskip

There are two types of evolution leading to formation of vortical flow in $2D
$ fluid, including $2D$ approximations of the planetary atmosphere and
magnetized plasmas. The first is the projection of the intrinsically
three-dimensional evolution and consists (for the atmosphere) of the large
scale instability involving a wide range of thermal processes:
buoyancy-induced convection, exchange of heat, phase transitions
(condensation, evaporation), etc. In this evolution vorticity is created
and convected by momentum fluxes and the saturation is reached at the balance
of sources and sinks.

A second type of evolution consists of the separation of opposite-sign
vorticities in different regions of the plane, together with concentration
of the like-sign vorticity. It acts on the existing vorticity, which usually
is randomly distributed in plane at the initial stage, with exact
conservation of the total vorticities of each sign: no creation and no
destruction. This is the \emph{self-organization of the vorticity}, leading
asymptotically to a highly coherent pattern of flow. This process does not
need any of the components involved in the first case: no temperature, no
pressure gradient, no buoyancy, no exchange of heat or phase transitions,
etc. It is just the spontaneous reorganization of the vorticity initially
present in the field, a well known property of fluids in two-dimensions. The
nature of the process is analogue to the Widom Rawlinson phase transition.

In real life these two processes take place simultaneously: vorticity is
created in the mechanical and thermodynamical processes (cyclogenesis), is
convected and is redistributed spatially through the velocity field that
results from the effects of forces and sinks. In the same time it takes
place the process of self-organization of vorticity, consisting of
exclusively interaction and merging of elements of vorticity.

The following question can be formulated: how much of the dynamics and of
the properties of the stationary state of an atmospheric vortex (in
particular a tropical cyclone) is determined by the process of
self-organization of vorticity ? There are two possible attitudes in
connection with this problem. One may be tempted to assume that the full set
of mechanical and thermodynamical processes simply suppresses the
manifestation of the spontaneous vorticity organization, overwhelmed by
intensely active processes. There is no proof however that this is so. Or,
one can assume that the spontaneous self-organization is in any case
embedded into the full framework and there is no reason to take care of it
in a particular way. That this is not a correct answer can be seen after the
most elementary tentative to describe the self-organization of the
vorticity: the methods are radically different of those used in cyclogenesis
and there is no sign that they would be embedded in the usual treatments.
Simple intuition may fail dramatically.

\bigskip

The fact that all tropical cyclones at stationarity have velocity fields
with qualitatively the same radial profile suggests that there may be a
connection with universal coherent structures (vortices) found in many other
systems: ideal fluid, superconductors, topological field theory, cosmic
matter, etc. In such systems the vorticity field evolves by
self-organization to states that extermize a functional, \emph{i.e.} they
are exceptional within the much wider class of functions that verify the
conservation equations for the same system. In the case of the $2D$ Euler
(non-dissipative, incompressible) fluid there is \textit{no} thermodynamic
process (as mentioned before, no buoyancy, no pressure gradient, no exchange
of heat) but the asymptotic organization of flow into coherent structures is
a well known and well studied fact \citep{KraichnanMontgomery}. The results
of \citet{Montgomery1992} and (\citeyear{Montgomery1993}), showing the
evolution of the $2D$ fluid from an initial turbulent state to a highly
organized vortical motion, are fully convincing. Deep in the tropical
cyclone dynamics there must be present the tendency of self-organization of
the vorticity, similar to the one in the case of the Euler fluid. The
description of the tropical cyclone must somehow include this spontaneous
self-organization and respond to questions like: \textquotedblleft is the
self-organization of vorticity the dominant factor, or is-it quantitatively
insignificant?\textquotedblright ; \textquotedblleft how the specific
description of this process [which is variational and cannot rely on only
conservation laws] is intertwined with the description of the thermal
processes, for which conservation laws are used?\textquotedblright . It may
result that the self-organization of vorticity is weak and slow and requires
too much time, etc. Alternatively, it may result that the asymptotic
stationary state of the tropical cyclone is dominated by the structure
emerging from self-organization of the vorticity.

Trying to answer these questions one immediately finds that the inclusion of
the self-organization of vorticity field into the theory of cyclogenesis is
very difficult.

The cyclogenesis works with conservation equations (density, momentum,
angular momentum, energy and phase transitions).

The self-organization of the vorticity field needs completely different
methods. The temperature, the density, etc. play no role, the process is
purely kinematic. Therefore the problem was to give a formalism for the
vorticity self-organization, before any attempt to merge this process with
the cyclogenesis.

\bigskip

At first sight there are few chances for the self-organization of vorticity
to have a significant effect on the characteristics of the tropical cyclone
at stationarity. Two essential requests seem very difficult to be satisfied:
(1) the self-organization needs two-dimensionality, while the tropical
cyclone cannot be reduced to $2D$; and (2) the thermodynamic processes will
always be very active - even at stationarity - and the supposedly weak and
slow self-organization of vorticity would be hidden by the dominant effect
of forces and sinks.

There are however regimes where both restrictions may be inefficient and the
self-organization of the vorticity can manifest itself as the dominant
factor. They are characterized by: the possibility (adequacy) of the
two-dimensional approximation for the tropical cyclone; and the weak
coupling between the balanced thermal processes and the mechanical processes
in this asymptotic state. The flows of the tropical cyclone are three
dimensional but with substantial anisotropy: the azimuthal flow is largely
dominant compared with the radial and the vertical flows. Experiments
clearly show $2D$ vorticity concentration in water tank experiments although
the flows are three dimensional \citep{HopfingerVanHeijst}. In numerical
simulation of the  turbulence of the planetary atmosphere the $2D$ vorticity
concentration has been observed \citep{McWilliamsWeissYavneh}, with clear
connection with the Taylor - Proudman theorem \citep{Batchelor}. Regarding
the other element mentioned above, one may expect that close to the
stationary state and assuming that the vortical structure as a whole is not
acted upon by external factors, the thermodynamical processes and the
mechanical balance are weakly-coupled. In this limit there is only a small
amount of energy flowing from the thermal sub-system toward the mechanical
processes, the amount needed for the latter to overcome the loss due to the
friction. The loss of the mechanical energy by friction in the vortical
motion is a small fraction of the total mechanical energy.

How useful is such approximation that factorizes the physical system at
stationarity into thermal and vorticity-dynamics subsystems? For such ideal
state one assumes that the thermal processes are balanced and places
emphasis on the vorticity dynamics, seen as the essential factor in
establishing the spatio-temporal characteristics of the atmospheric vortex
at stationarity (the "shape"). This opens the possibility that the
self-organization of the vorticity can manifest itself as the dominant
process in the asymptotic (quasi-stationary) state of the tropical cyclone:
the two-dimensionality and the self-organization of the vorticity are
strongly connected. In addition, it may also act freely if the thermal
processes are almost balanced. If indeed there is an universal vortical
structure behind the stationary tropical cyclone then this would only be the
result of its dominant two-dimensional geometry and of the free
manifestation of the self-organization of vorticity.

\bigskip

It is not possible to discuss here the vast analytical and numerical effort
dedicated to understanding the self-organization of $2D$ ideal fluid's
vorticity. Few comments are however necessary in preparation of our
presentation below.

The ideal (Euler) fluid is described in $2D$ by the equation $d\mathbf{%
\omega /}dt=0$ where $\mathbf{\omega }$ is the vorticity, a vector directed
along the perpendicular on the plane. This equation is known (since works of
Kirchhoff and Helmholtz) to be equivalent with the equations of motion of a
discrete set of point-like vortices interacting in plane by a long-range
potential (Coulombian, the logarithm of the relative distance between
point-like vortices). This system has been treated as a statistical ensemble
(\citet{Onsager}, \citet{KraichnanMontgomery}, \citet{EdwardsTaylor}) with
finite phase space. The statistical temperature is \emph{negative} for any
positive value of the energy. The extremum of the entropy, under the
constraints of fixed energy and fixed, equal, numbers of positive and of
negative vortices, has led to the \emph{sinh}-Poisson equation for the
streamfunction of the flow, which was later confirmed by numerical
simulations \citep{Montgomery1992}. Several works, attempting extension of
this result but remaining in the same statistical approach have produced
different equations for the asymptotic flows ( \citet{Pasmanter}, \citet%
{LundgrenPointin}; for a review see \citet{Chavanis2}). Other studies have
focused on the dynamics of few point-like vortices (\citet{Aref1983}, \citet%
{Novikov1975}, \citet{Newton}, \citet{MajdaBertozzi}). Besides the interesting
aspect of integrability they can be applied when the flow is potential on
most of the domain, these studies being therefore relevant to superfluidity.

A related approach starts from the Kelvin circulation theorem and divides
the vorticity initially present in the field in patches of finite extension
(\textquotedblleft vortex patches\textquotedblright , \citet{Saffman}, \citet%
{Aref1983}, \citet%
{GustafsonSethian}), following their dynamics as convected by the flow \citep%
{RobertSommeria1992}. These methods have been used by Holland and
collaborators (\citet{LanderHolland}, \citet{RitchieHolland}, \citet%
{HollandDietachmayer}, \citet{WangHolland}) to study the interaction and
merging of vortices in connection with the generation of the tropical
cyclone.

\bigskip

The approach that we have developed (and is used in the present work)
consists of the formulation of the continuum limit of the discrete set of
point-like vortices in terms of a classical field theory \citep%
{FlorinMadi2003, FlorinMadi2005, FlorinMadiGAFD}. The
evolution of the $2D$ ideal Euler fluid to vorticity organization is
governed by the extremum of an action functional. The asymptotic states are
stationary, have the property of \textquotedblleft
self-duality\textquotedblright\ and satisfy the equation $\sinh $ - Poisson
equation (also known as elliptic sinh - Gordon equation) $\Delta \psi +\sinh
\left( \psi \right) =0$, where $\psi $ is the streamfunction. The
self-duality (SD) is a property of the geometric - algebraic structure (a
fiber space) attached to the physical problem: the curvature differential
two - form is equal to its Hodge dual \citep{MasonWoodhouse}. Identification
of this mathematical structure is highly non-trivial but in practical cases
SD is manifested by the possibility of expressing the action functional as a
sum of square terms plus a term with topological content. The $\sinh $ -
Poisson equation has been derived from an action that has the SD property.
The equation is exactly integrable and the doubly periodic solutions
represent the absolute minimum of the action.

It is not our intention to contrast the \textquotedblleft
statistical\textquotedblright , the \textquotedblleft vortex
patches\textquotedblright\ and the \textquotedblleft
field-theoretical\textquotedblright\ approaches, even less in the present
work, which has a different subject. Comparisons can still be made, \citep%
{FlorinMadiXXX2013} hampered by the very different theoretical formulations.

\bigskip

In the case of the $2D$ model for the atmosphere the $\sinh $ - Poisson
equation cannot be more than an indicative approximation. This is because
there is a new physical element, the Rossby radius, that changes the physics
and the mathematical possibility of relaxed states.

For the $2D$ approximation of the planetary atmosphere (and for the $2D$
plasma in strong magnetic field: the equations are the same) the dynamics of
the vorticity field can be equivalently described by a discrete system of
point-like vortices (\textquotedblleft geostrophic point
vortex\textquotedblright\ according to Morikawa \citeyear{Morikawa1960}) but in
this case the potential of interaction in plane is short - range.
 The continuum limit of the system of discrete point-like
vortices is again a classical field theory. The matter field $\phi $ (which
represents the density of the point-like vortices) and the gauge field
(representing the mutual interaction of the vortices) are elements of the
algebra $sl\left( 2,\mathbf{C}\right) $, \emph{i.e.} they are mixed spinors,
since they correspond to physical elementary \emph{vortices}. The planetary
rotation represents the \textquotedblleft condensate of matter\textquotedblright\ that defines the broken
vacuum of the theory and generates, via the Higgs mechanism, the mass of the
\textquotedblleft photon\textquotedblright\ , \emph{i.e. }the short range of the interaction, with the spatial
decay given by the Rossby radius (respectively the Larmor gyro-radius for
plasma). In the following we just remind few elements of the Field
Theoretical (FT) formulation for the $2D$ atmosphere/plasma. The FT
formulation can be found in \citep{FlorinMadi2005}, \citep{FlorinMadiXXX}
and the first application in \citep{FlorinMadiGAFD}.

The Lagrangian density is 
\begin{eqnarray}
\mathcal{L} &=&-\kappa \varepsilon ^{\mu \nu \rho }\mathrm{tr}\left(
\partial _{\mu }A_{\nu }A_{\rho }+\frac{2}{3}A_{\mu }A_{\nu }A_{\rho }\right)
\label{26} \\
&&-\mathrm{tr}\left[ \left( D^{\mu }\phi \right) ^{\dagger }\left( D_{\mu
}\phi \right) \right]  \nonumber \\
&&-V\left( \phi ,\phi ^{\dagger }\right)  \nonumber
\end{eqnarray}

where $\kappa $ is a positive constant and%
\begin{equation}
V\left( \phi ,\phi ^{\dagger }\right) =\frac{1}{4\kappa ^{2}}\mathrm{tr}%
\left[ \left( \left[ \left[ \phi ,\phi ^{\dagger }\right] ,\phi \right]
-v^{2}\phi \right) ^{\dagger }\left( \left[ \left[ \phi ,\phi ^{\dagger }%
\right] ,\phi \right] -v^{2}\phi \right) \right] .  \label{vpot}
\end{equation}

The field variables are $\phi $, $A^{\mu }\equiv \left(
A^{0},A^{1},A^{2}\right) $ and their Hermitian conjugate, $\left( {}\right)
^{\dagger }$. The covariant derivative is $D_{\mu }=\partial _{\mu }+\left[
A_{\mu },\right] $, $\mu =0,1,2$ and the metric $g^{00}=-1$, $g^{ik}=\delta
^{ik}$. All variables are elements of the algebra $sl\left( 2,\mathbf{C}%
\right)$. A standard Bogomolnyi procedure followed by an algebraic ansatz
where $\phi $ only contains the two ladder generators of $sl\left( 2,\mathbf{%
C}\right)$ leads to an equation for the asymptotic states that has no
regular real solution. Adopting an algebraic ansatz with only the first
ladder generator in $\phi $ leads to a very clear topological theory but the
asymptotic equation can only produce stationary rings of vorticity. If we
see the field theoretical description of the atmospheric vortex as an
extension of the theory for the Euler fluid, then we have to keep the
Bogomolnyi procedure, but alter the terms: the action functional becomes as
usual a sum of squares plus a residual term. This term is small (being
multiplied with the Coriolis frequency $\sim 5 \times 10^{-5}$) and does not
have a topological meaning \citep{FlorinMadiXXX}. The self-duality property
is not exact but the resulting equation \citep{FlorinMadi2005}%
\begin{equation}
\Delta \psi +\left( \frac{v^{2}}{\kappa }\right) ^{2}\sinh \left( \psi
\right) \left[ \cosh \left( \psi \right) -1\right] =0  \label{eq}
\end{equation}%
has solutions with the morphology of the tropical cyclone. With the
identifications%
\begin{eqnarray}  \label{v2f}
v^{2} &=&f \\
\kappa &=&\sqrt{gh_{0}}  \label{kappagh}
\end{eqnarray}%
we see that the distances should be normalized to the Rossby radius $%
R_{Rossby}=\sqrt{gh_{0}}/f$ and the time to $f^{-1}$, the inverse of the
Coriolis frequency. The equation is solved on: (A) a square with half of the
length of diagonal $L^{diag}$ (we will also use $L^{sq}=L^{diag}/\sqrt{2}$,
half of the length of the side ); and (B) in azimuthal symmetry on a radial
interval $L^{rad}$. The results coincide for $L^{rad}=L^{diag}=\sqrt{2}%
L^{sq} $, as described in \citep{FlorinMadiGAFD}. From now on the quantities
like $L^{diag}$, $L^{rad}$, $L^{sq}$, $\psi $, etc. are normalized and when
they are dimensional an upperscript $phys$ is used: $\left( L^{diag}\right)
^{phys}=R_{Rossby}L^{diag}$, etc. We note that here $R_{Rossby}$ is defined
as a global physical parameter of the tropical cyclone and we do not
consider either its spatial variation within a single vortex or the $\beta$
effect.

Our simplified model for the tropical cyclone now can be formulated in the
terms of the two approaches (geophysical and field theoretical). In the
present work we underline a result that is derived in the field theoretical
description and reveals a strong property in the geophysical picture of the
tropical cyclone: the field theoretical result that the mass of the matter
field excitation $m_{H}$ is equal with the mass of the gauge boson $%
m_{gauge} $ implies that the maximum radial extension of the tropical
cyclone must be equal to the Rossby radius.

\bigskip

This is an important and strongly constraining condition on the physical
dimensions of a tropical cyclone. According to the simplified, field
theoretical (FT) model, if the physical dimensions are so different for
different tropical cyclones, this is due to different Rossby radii. Or, the
Rossby radius results from the individual history of a particular tropical
cyclone, which, after the transient part of growth, should reach a unique
shape, given by the solution of the Eq.(\ref{eq}) for $L^{rad}=1$. In the FT
framework the property $m_{H}=m_{gauge}\sim \left( L^{rad}\right) ^{-1}$
means that $L^{rad}=\left( L^{rad}\right) ^{phys}/R_{Rossby}=1$. This gives
a \emph{unique} profile $\psi \left( r\right) $ which is obtained by solving
the Eq.(\ref{eq}) either on a square region in the plane, or on a radial
domain $L^{rad}=1$. We remind that the result of solving Eq.(\ref{eq}) is
expressed in non-dimensional quantities: distances are normalized to $%
R_{Rossby}$ and velocity to $\left( R_{Rossby}f\right) $. A physical input
coming from observations is necessary to get dimensional quantities.
Analyzing a database we can find $R_{Rossby}$ for a particular atmospheric
vortex and then calculate the maximal velocity, radius of eye-wall, etc.,
which must be compared with observational data.

We are interested in three important characteristics of the cyclone: the
maximum of the azimuthal velocity $v_{\theta }^{\max }$, the radius where
this maximum is found $r_{v_{\theta }^{\max }}$ and the maximal radial
extension of the vortex, $R_{\max }$. The use of observational data to
identify the Rossby radius and then to convert our variables to physical
ones, followed by further comparisons, is however a difficult task: our
simple model refers to the stationary state of the tropical cyclone, which
is difficult to isolate in the full evolution. Second, when two of the three
characteristics mentioned above are fixed, the dispersion of observational
data regarding the third one is large. We associate this dispersion with the
fact that the state of the tropical cyclone cannot be exactly mapped to the
vortex derived in the field theoretical formulation at self-duality and its
shape does not correspond to $L^{rad}=1$. We then use a range of values
around $L^{rad}=1$ and try to find the effective $L^{rad}\in \left[ L_{\min
}^{rad},L_{\max }^{rad}\right] $ which provides the best fit to the measured
data. This means that we assume that the system evolves in close proximity
of the stationary Self-Dual state. In short the FT leads us to expect that
for whatever physical dimensions of the tropical cyclones, we should find $%
L^{rad}=1$. If we find a different value this means that the special state
of self-duality, leading to Eq.(\ref{eq}) is not reached and $\left(
L^{rad}\right) ^{phys}\approx $ $R_{Rossby}$ is not fulfilled. We would
like to see to what extent the FT remains an interesting description in the
neighborhood of this particular state.

\section{The geophysical view on the typical dimensions of the atmospheric
vortex}

For the $2D$ model of the atmosphere the potential of interaction between
the discrete point-like vortices \citep{Morikawa1960} is no more long range
(Coulombian $\ln \left( \left\vert \mathbf{r-r}^{\prime }\right\vert \right) 
$), it is $K_{0}\left( \sigma \left\vert \mathbf{r}-\mathbf{r}^{\prime
}\right\vert \right) $, with $\sigma ^{2}=f^{2}/\left( gh_{0}\right) $. Here 
$f$ is the Coriolis frequency $f=2\Omega \sin \theta $, $\Omega $ is the
frequency of planetary rotation and $\theta $ is the latitude angle; $g$ is
the gravitational acceleration and $h_{0}$ is the depth of the fluid
(atmosphere) layer. The space parameter is defined \citep{Pedlosky} as the
Rossby radius of deformation. 
\begin{equation}
R_{Rossby}=\frac{\left( gh_{0}\right) ^{1/2}}{f}=\sigma ^{-1}
\label{Rossbydef}
\end{equation}

Besides $R_{Rossby}$ there is another natural space parameter, $L$, the
characteristic horizontal length $L$ of the flow induced by a perturbation
of the atmosphere ($L$ is dimensional). These two parameters control the
balance of the forces in the fluid dynamics. The relative acceleration of
the flow $d\mathbf{u}/dt$ results from a competition of the forces induced
by the horizontal gradient of the pressure $-\mathbf{\nabla }_{h}p$ and the
Coriolis force $\mathbf{u\times }2\mathbf{\Omega }$. The gradient of
pressure exists due to the perturbation of the pressure of the air $p=\rho
gz $, created by the perturbation of the \emph{depth} $z=h_{0}+\delta z$ of
the layer of the fluid,%
\begin{equation}
-\frac{\partial }{\partial x}\delta p=-\rho _{0}g\frac{\partial }{\partial x}%
\delta z\sim -\rho _{0}g\frac{\delta z}{L}  \label{pexp}
\end{equation}

We note that, for a perturbation $\delta z$ of the depth of the fluid layer,
if the horizontal extension of the flow $L$ is large, the gradients $%
-\partial p/\partial x$ , $-\partial p/\partial y$ are small ($\sim \delta
z/L$) and the Coriolis force is dominant. This term, proportional with $%
\delta z\sim \psi \left( x,y\right) $ leads to the second part of the \emph{%
potential vorticity} 
\begin{equation}
\Pi \equiv \mathbf{\nabla }_{h}^{2}\psi -\sigma ^{2}\psi =\mathbf{\nabla }%
_{h}^{2}\psi -\frac{1}{R_{Rossby}^{2}}\psi  \label{potvort}
\end{equation}%
In the geostrophic approximation $\Pi $ verifies the conservation equation $%
d\Pi /dt=0$, where the convective derivative operator is $d/dt=\partial
/\partial t+u\partial /\partial x+v\partial /\partial y$. The velocity $%
\mathbf{v=}\left( u,v\right) $ is defined in terms of the streamfunction $%
\psi \left( x,y\right) $, $\mathbf{v=}-\mathbf{\nabla }_{h}\psi \times 
\widehat{\mathbf{e}}_{z}$, with $\widehat{\mathbf{e}}_{z}$ the versor
perpendicular on the plane and $\mathbf{\nabla }_{h}$ is the horizontal
gradient. The relative vorticity $\mathbf{\nabla }_{\perp }^{2}\psi $
introduces the horizontal scale of the flow, $\mathbf{\nabla }_{h}^{2}\sim
L^{-2}$; the contribution to the potential vorticity of the deformation of
the free surface ($\delta z$, the perturbed height of the fluid layer)
introduces the Rossby radius $R_{Rossby}=\sqrt{gh_{0}}/f$. The importance of
the term coming from the deformation of the surface, $\sim \psi $, relative
to the vorticity term $\mathbf{\nabla }_{\perp }^{2}\psi $ is measured by
the factor \citep{Pedlosky}%
\begin{equation}
F=\left( \frac{L}{R_{Rossby}}\right) ^{2}  \label{factor}
\end{equation}%
and two regimes are identified. (1) If the horizontal scale $L$ is small and
localized inside the Rossby radius scale%
\begin{equation}
L\ll R_{Rossby}  \label{case1LR}
\end{equation}%
then from the point of view of the vorticity balance the free surface can be
considered flat \ and rigid (\emph{i.e.} no deformation). In relative terms,
a very large Rossby radius means that the external origin of rotation is
weak (in the equivalent plasma system, a very large Larmor radius means that
the applied external magnetic field is weak). The operator $\Delta _{h}\psi
-R_{Rossby}^{-2}\psi $ approaches $\Delta _{h}\psi $ and the short range
interaction in the system of point-like vortices turns into the long range
interaction, $K_{0}\rightarrow \ln $. The density and the vorticity decouple
and the Ertel's theorem becomes the simple statement of conservation of the
vorticity $d\Delta _{h}\psi /dt=0$, \emph{i.e.} the Euler equation. (2) If
the horizontal extension of the perturbation flow is very large, much larger
than the Rossby radius 
\begin{equation}
L\gg R_{Rossby}  \label{case2LR}
\end{equation}%
the \emph{relative vorticity} in the motion $\Delta _{h}\psi $ is very small
and the velocity field appears almost uniform horizontally. For large
spatial scales of the flow $L$, the \emph{relative accelerations} are very
weak and the Coriolis acceleration dominates.

Then the basic geophysical analysis finds the Rossby radius of deformation $%
R_{Rossby}\approx L$ as the "distance over which the gravitational tendency
to render the free surface flat is balanced by the tendency of the Coriolis
acceleration to deform the surface" \citep{Pedlosky}.

The fact that the horizontal extension of the tropical cyclone is comparable
with the Rossby radius has been noted before \citep{Willoughby1}.

\bigskip

\section{Field theoretical view on the typical (radial) dimensions}

In FT the spatial decay of the interaction is connected with the mass of the
particle that carries the interaction. The FT formulation of the atmospheric
vortex allows to consider, instead of the typical lengths $L$ and $%
R_{Rossby} $, the masses associated with the propagators of the scalar and
gauge fields excitations.

In field theory formalism the mass appears as a singularity of the
propagator of the field, which is calculated as the two-point correlation of
the field values. Alternatively, to identify the mass $m_{H}$ of the scalar $%
\phi $ field excitation, we need to emphasize from the equations of motion
derived from the Lagrangian, a structure expressing the main scalar field
dynamics, as%
\begin{equation}
-\partial _{i}^{2}\phi -\left( m_{H}\right) ^{2}\phi  \label{oscillator}
\end{equation}%
and this can be seen in the expression of the action functional, without the
need to calculate the propagator \citep{AbersLee}. The second order
differential operator comes from the kinetic term in the Lagrangian $\left(
D_{\mu }\phi \right) ^{\dagger }\left( D^{\mu }\phi \right) $ and the last
term comes from the part of the potential $V\left( \left\vert \phi
\right\vert ^{2}\right) $ which is quadratic in $\phi $. It is simpler to
refer to the Abelian version of the Lagrangian \citep{DunneBook} (in this
case we refer to the matter field $\phi $ as the "scalar" field). For the
Abelian version, instead of Eq.(\ref{vpot}) the potential is 
\citep[page
83]{DunneBook}%
\begin{equation}
V\left( \phi \right) =V\left( \left\vert \phi \right\vert ^{2}\right) =\frac{%
1}{4\kappa ^{2}}\left\vert \phi \right\vert ^{2}\left( \left\vert \phi
\right\vert ^{2}-v^{2}\right) ^{2}  \label{potab}
\end{equation}%
and this identifies the broken vacuum as%
\begin{equation}
\left\vert \phi _{0}\right\vert ^{2}=v^{2}  \label{brokvac}
\end{equation}%
In order to find the mass spectrum in the broken vacuum, we have to expand
the potential around $\left\vert \phi _{0}\right\vert ^{2}=v^{2}$ and retain
the quadratic terms like in Eq.(\ref{oscillator})%
\begin{eqnarray}
&&V\left( \phi _{0}+\widetilde{\phi }\right)  \nonumber \\
&=&\frac{1}{4\kappa ^{2}}\left\vert \phi _{0}\right\vert ^{4}\left( 
\widetilde{\phi }+\widetilde{\phi }^{\ast }\right) ^{2}+...  \label{vexp}
\end{eqnarray}%
The field $\widetilde{\phi }$ is complex $\widetilde{\phi }=\phi _{1}+i\phi
_{2}$ which gives $\widetilde{\phi }+\widetilde{\phi }^{\ast }=2\phi _{1}$
(where $\phi _{1}\equiv \mathrm{{Re}\left( \phi \right) }$) and replacing in
the expression of the expanded potential $V$ we have%
\begin{equation}
V\left( \phi _{0}+\widetilde{\phi }\right) =\frac{v^{4}}{\kappa ^{2}}\phi
_{1}^{2}+...  \label{vexp2}
\end{equation}%
There is a single \emph{real} field with mass 
\begin{equation}
m_{H}=\frac{v^{2}}{\kappa }  \label{massphi}
\end{equation}

\bigskip

For the gauge field, again taking the Abelian form for simplicity, the
following part in the Lagrangian, which can lead to the identification of a
mass for the gauge field, is%
\begin{equation}
-\kappa \varepsilon ^{\mu \nu \rho }\left( \partial _{\mu }A_{\nu }\right)
A_{\rho }-\left\vert \phi _{0}\right\vert ^{2}A_{\mu }A^{\mu }  \label{gaug}
\end{equation}%
The first term is the Chern - Simons term in the Lagrangian and the second
term comes from the square of the covariant derivative (in the kinetic term $%
\left( D_{\mu }\phi \right) ^{\dagger }\left( D^{\mu }\phi \right) $), after
taking the scalar field in the \emph{vacuum} state, $\phi \rightarrow \phi
_{0}$. This gives a mass%
\begin{equation}
m_{gauge}=\frac{1}{\kappa }\left\vert \phi _{0}\right\vert ^{2}=\frac{v^{2}}{%
\kappa }  \label{massgaug}
\end{equation}

It results 
\begin{equation}
m_{H}=m_{gauge}=\frac{v^{2}}{\kappa }  \label{equal}
\end{equation}

\bigskip

The identification of the mass spectrum of the field particles for the
action functional Eq.(\ref{26}) with (\ref{vpot}) is complicated by the
non-trivial algebraic content of the theory. As above, the masses of the
excitations around the broken vacuum $\phi _{0}$ are obtained by expanding $%
V\left( \phi _{0}+\widetilde{\phi }\right) $. As shown by Dunne \citeyear{DunneBook},
retaining the quadratic terms in the expansion leads to a matrix and the
mass spectrum is determined from the eigenvalues of this matrix. The
relationship between the masses of the scalar field (Higgs) particle and of
the gauge particle is the same $m_{H}=m_{gauge}= v^{2}/\kappa $ .

We \textbf{note} that the mass of the vector potential is related to the
condensate of the vorticity, which is the vacuum of the theory ($\left\vert
\phi _{0}\right\vert ^{2}=v^{2}$): the Coriolis frequency. This is the
background on which exists any perturbation of velocity/vorticity. Due to
the planetary rotation the interaction between two elements of vorticity in
the atmosphere decays \ on the length of the Rossby radius.

\section{Comments on the relationship between the two views on the
characteristic parameters of the atmospheric vortex}

The two descriptions refer to the same physical reality. In the geophysical
formulation the state where the two characteristic lengths are comparable, $%
L\approx R_{Rossby}$ has been identified as having particular properties. In
the field theoretical formulation the equality $m_{H}=m_{gauge}$ (which
through the mapping corresponds to $L=R_{Rossby}$) indicates a state with
exceptional properties, the self-duality. \ Now we should recall that the
fundamental property that is behind the high organization of the vorticity
in the Euler asymptotic states is the \emph{self-duality}, which is only
revealed by the FT formulation. It is an admitted fact that any coherent
structure known to date (solitons, instantons, topological field
configurations, etc.) owes its existence to the self-duality %
\citep{MasonWoodhouse}. Therefore, the well known experimental observation
of vorticity organization into coherent structure of the flow in the Euler
fluid naturally suggested to look for self-duality and the FT formulation
confirmed that indeed the SD exists.

In the case of the atmospheric vortex (as for $2D$ magnetized plasma) the SD
state is only an approximation but we are still led to follow the suggestion
that the existence of a quasi-stationary, quasi-coherent vortex like the
idealized tropical cyclone is due to this approximative self-duality. Then
the particular relationships: $L\approx R_{Rossby}$ and $m_{H}=m_{gauge}$
are associated to self-duality and the atmospheric vortex that verifies this
condition is quasi-coherent. This is the reason that the tropical cyclone
has the highest state of organization and the highest stability.

Solving the Eq.(\ref{eq}) for $L\neq 1$ means that we consider that the
departures from the self-duality state can still be reflected by the
Lagrangian dynamics and \ this can be obtained from the same equation but
for unbalanced lengths $L\neq R_{Rossby}$, which may be supposed that
reflects different masses for the matter and gauge fields.~We do not have a
demonstration for this. We just note that this point of view is similar to
the procedure adopted by \citep{JacobsRebbi} to calculate the energy of
interaction of an ensemble of Abrikosov Nielsen Olesen vortices in
superfluids in close proximity of the self-dual state; also, it is similar
to the assumption adopted by \citep{Manton} in the calculation of the motion
of the vortices, near self - duality, as geodesic motion on the manifold
consisting of the moduli space of a set of vortices which are solutions of
the Abelian-Higgs model.

\bigskip

\section{Numerical studies close to the equality of the two radial lengths}

\subsection{The relationships between the main characteristics of the
tropical cyclone, derived in the Field Theoretical approach}

We have constructed a field theoretical model of the dynamics of the
point-like vortices in plane and on this basis we study the
self-organization of vorticity in a $2D$ approximation of the atmosphere. By
no means we cannot claim that we cover the full complexity of the real
tropical cyclone: our description can approach the reality in certain
restrictive cases: the $2D$ approximation is acceptable, the stationarity is
ensured, the vorticity self-organization is dominant compared to thermal
processes. Our expectations can be formulated in this way: if the comparison
between our quantitative results and the observations is favorable, it means
that the self-organization of the vorticity is a substantial part of the
dynamics and that our FT model is adequate.

\bigskip

Using a large number of solutions of Eq.(\ref{eq}) we have identified
systematic relationships between the three characteristics, with only the
parameter $L^{rad}$ \citep{FlorinMadiGAFD}. The differential equation has
been solved both on a plane square and on the radius in cylindrical
symmetry, for an interval of $L^{rad}=\sqrt{2}L^{sq}$ around $1$. The
results allow to find two relationships between the tropical cyclone
parameters: the radius of the circle where the azimuthal velocity is
maximum, $r_{v_{\theta }^{\max }}$, the magnitude of the maximum of the
azimuthal velocity $v_{\theta }^{\max }$ and the maximum radial extension of
the cyclone, $R_{\max }$.%
\begin{equation}
v_{\theta }^{\max }\left( L^{sq}\right) =2.6461\times \exp \left( \frac{1}{%
L^{sq}}\right) -2.7748  \label{vmax}
\end{equation}%
A simple approximation \ is%
\begin{equation}
v_{\theta }^{\max }\left( L^{sq}\right) \approx e\left[ \exp \left( \frac{1}{%
L^{sq}}\right) -1\right]  \label{vmaxs}
\end{equation}%
where $e\equiv \exp \left( 1\right) $.

\begin{figure}[tbph]
\centerline{\includegraphics[height=5cm]{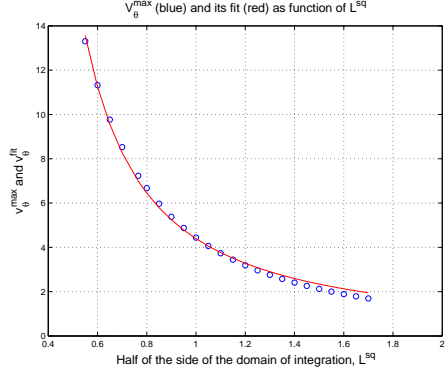}}
\caption{The analytical fit of the maximum velocity resulting from solving the Eq.(3).}
\label{fig1}
\end{figure}

\begin{figure}[tbph]
\centerline{\includegraphics[height=5cm]{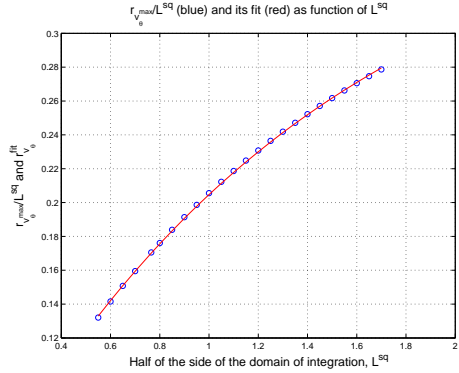}}
\caption{The analytical fit of the ratio: radius where the maximum velocity is found
over the length $L^{sq}$. This is inferred from a large set of solutions of
the eq.(3) for various space domains $\left( L^{sq}\text{\ \ or}\ \ L^{rad}=%
\sqrt{2}L^{sq}\right) $.}
\label{fig2}
\end{figure}

The other relation is%
\begin{equation}
\frac{r_{v_{\theta }^{\max }}}{L^{sq}}\left( L^{sq}\right) =0.395892+0.386360%
\left[ -\exp \left( -\frac{L^{sq}}{\sqrt{2}}\right) \right]  \label{rvmax}
\end{equation}%
with a simple approximation%
\begin{equation}
\frac{r_{v_{\theta }^{\max }}}{\sqrt{2}L^{sq}}\approx \frac{1}{4}\left[
1-\exp \left( -\frac{\sqrt{2}L^{sq}}{2}\right) \right]  \label{rvmaxs}
\end{equation}

[Note that, compared with a previous work \citep{FlorinMadiGAFD}, we have
eliminated the factor $1/2 $ in front of the nonlinear term in Eq.(\ref{eq}%
). In general an arbitrary factor $\lambda $ can be used as long as the
scaling of the coordinates is made $x^{\prime }=x/\sqrt{\lambda }$, but in
the present case taking $\lambda =1$ makes easier the comparison with
observations. Another difference is a better procedure of fit that have led
to an improved calculation of the coefficients in the two equations above].

The two relationships Eqs.(\ref{vmax}) and (\ref{rvmax}) will first be used
in conjunction with the observational data which we take from the paper of
Shea and Gray (\citeyear{SG}). The objective is to examine consequences of the
relationship discussed in this work: $L^{rad}\approx 1$, or $\left(
L^{rad}\right) ^{phys}\approx R_{Rossby}$. We expect to find a
clusterization of observational data around those results that take into
account this relationship.

\bigskip

\subsection{Procedure to obtain physical data from the FT equations with
input from observations}

\subsubsection{Calculation of the Rossby radius using input from Shea - Gray
database}

Shea and Gray \citeyear{SG} have organized a large set of observations in a
graphical representation of the relationship between the radius where the
maximum azimuthal velocity is measured and the magnitude of the maximum
velocity, in our notations $\left( r_{v_{\theta }^{\max }},v_{\theta }^{\max
}\right) ^{phys}$. This is Fig.45 of their paper. A line represents the best
fit and we will refer to its points as \textquotedblleft
SG\textquotedblright\ data in the following. The figure also shows a
substantial dispersion of the observed points. The best-fit line limits the
maximum velocity that we can use for comparisons to a range between $70\
knots$ and $115\ knots$ ($36$ to $59\ m/s$). Assuming that the set of points
of the fitting line in SG is parameterized by $R_{Rossby}$ we find for each
pair $\left( r_{v_{\theta }^{\max }},v_{\theta }^{\max }\right) ^{phys}$ the
corresponding $R_{Rossby}$ using the following procedure.

We start by taking a value of the normalized radius from the range $%
r_{v_{\theta }^{\max }}\in \left[ 0.1,0.25\right] $ and solve Eq.(\ref{rvmax}%
) for $\left( L^{sq}\right) $. It results $L^{sq}\in \left[ 0.655,1.13\right]
$. Each $\left( L^{sq}\right) $ is then inserted into the Eq.(\ref{vmax})
and the resulting velocity $\left( v_{\theta }^{\max }\right) $ is compared
with the data from SG. For this we need to return to dimensional variables 
\emph{i.e.} to multiply with $R_{Rossby}f$ , $v_{\theta }^{\max }\rightarrow
\left( v_{\theta }^{\max }\right) ^{phys}$. Since at this point $R_{Rossby}$
is not known we start with an initial value and solve iteratively until the
equality is obtained%
\begin{equation}
\left( v_{\theta }^{\max }\right) ^{phys}-v^{SG}=0  \label{eqvsg}
\end{equation}%
This equation for $R_{Rossby}$ leads to $R_{Rossby}\ \left( meters\right)
\in \left[ 106\times 10^{3},190\times 10^{3}\right] $

\begin{figure}[tbph]
\centerline{\includegraphics[height=5cm]{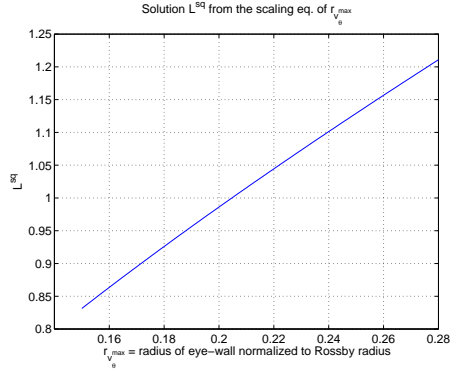}}
\caption{The length of the half-side of the square domain of integration in plane $%
L^{sq}$, for a fixed value of $r_{v_{\theta }^{\max }}$ . Both are
non-dimensional (normalized to $R_{Rossby}$).}
\label{fig3}
\end{figure}

\begin{figure}[tbph]
\centerline{\includegraphics[height=5cm]{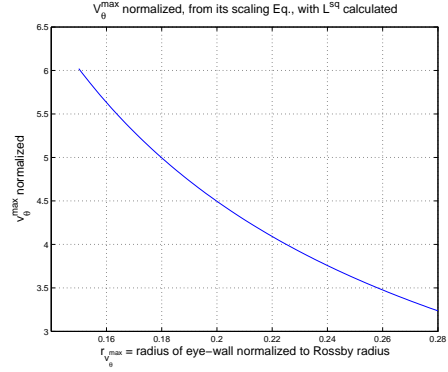}}
\caption{The maximum velocity $v_{\theta }^{\max }$ as results form the analytical
formula. The parameter $L^{sq}$ is determined previously for given values of 
$r_{v_{\theta }^{\max }}$ (see Fig.3).}
\label{fig4}
\end{figure}

\begin{figure}[tbph]
\centerline{\includegraphics[height=5cm]{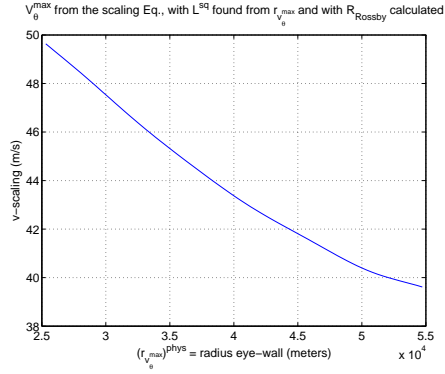}}
\caption{The maximum velocity $v_{\theta }^{\max }$ (dimansional, $m/s$) after the
Rossby radius is calculated using input from the Shea Gray data).}
\label{fig5}
\end{figure}

\begin{figure}[tbph]
\centerline{\includegraphics[height=5cm]{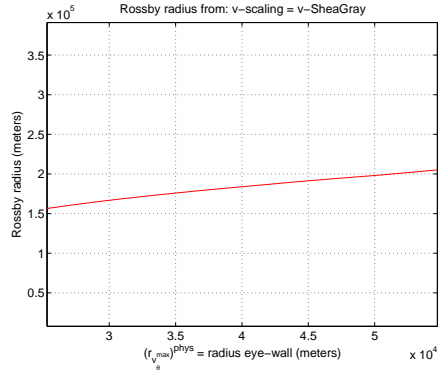}}
\caption{The Rossby radius as results form imposing equality of the maximum
velocities: from the analytical formula and from Shea Gray data.}
\label{fig6}
\end{figure}

\begin{figure}[tbph]
\centerline{\includegraphics[height=5cm]{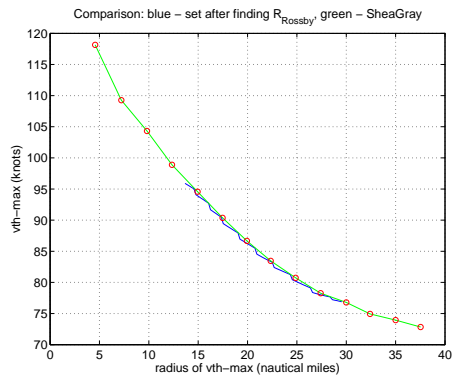}}
\caption{The range of Shea Gray data effectively used: the green line represents all
SG data and the blue line is the sub - domain used for calculation of the
Rossby radius.}
\label{fig7}
\end{figure}

\begin{figure}[tbph]
\centerline{\includegraphics[height=5cm]{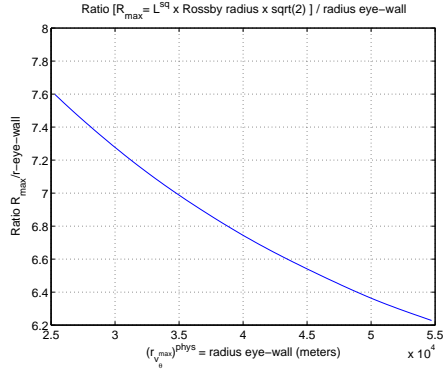}}
\caption{The ratio of the maximum radial extension ($L^{sq}\sqrt{2}\times R_{Rossby}$%
) and the radius of the maximum azimuthal velocity. This is in general a
number substantially greater than unity. After the determination of $%
R_{Rossby}$ we find a range $\left[ 6...8\right] $.}
\label{fig8}
\end{figure}

\begin{figure}[tbph]
\centerline{\includegraphics[height=5cm]{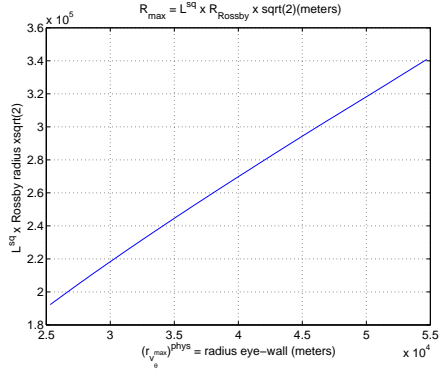}}
\caption{The maximum radial extension $R_{\max }=L^{sq}\sqrt{2}\times R_{Rossby}$ $%
\left( m\right) $ versus the radius of the eye-wall $\left( r_{v_{\theta
}^{\max }}\right) ^{phys}$ expressed in maters.}
\label{fig9}
\end{figure}

\begin{figure}[tbph]
\centerline{\includegraphics[height=5cm]{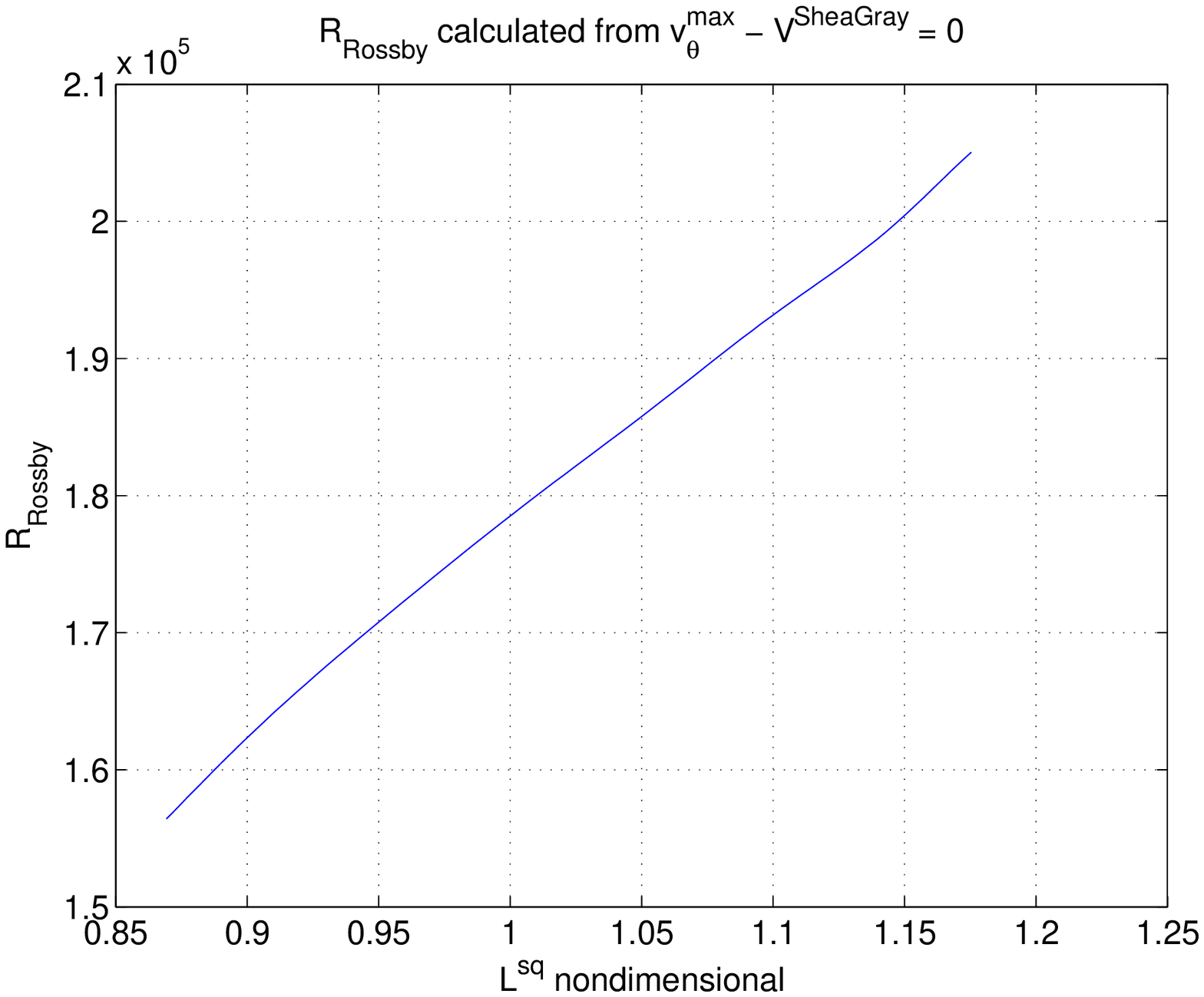}}
\caption{The Rossby radius obtained by imposing the equality of the velocity as given
by the analytical fit with the velocity from the Shea Gray data.}
\label{fig10}
\end{figure}

Using the SG data we can obtain a qualitative image of the relationships
between the physical parameters $\left( v_{\theta }^{\max },r_{v_{\theta
}^{\max }}\right) ^{phys}$ and $R_{Rossby}$ with the maximal radial
extension $\left( L^{rad}\right) ^{phys}=L^{sq}\sqrt{2}\times R_{Rossby}$
(already included in Eq.(\ref{rvmax})).

\bigskip

As shown in the previous work \citep{FlorinMadiGAFD} the two equations (\ref%
{vmax}) and (\ref{rvmax}) are able to correctly reproduce physical
characteristics of the tropical cyclone when the physical input is close to
the stationary state, which is the only state that can be described by the
Eq.(\ref{eq}). The radial profile of the azimuthal velocity $v_{\theta
}\left( r\right) $ obtained from integration of Eq.(\ref{eq}) also
reproduces the Holland empirical formula, for the cases where data are
available (Fig.10 of \citep{FlorinMadiGAFD}). To obtain a more general (even
if approximative) idea about the ability of calculated $v_{\theta }\left(
r\right) $ to reproduce observed profiles we have compared a large set of
radial integration results for various $L^{rad}$ with the empirical formulas,
like, $v_{\theta }^{HB}\left( r\right) =v_{\theta }^{\max }\left(
r_{v_{\theta }^{\max }}/r\right) ^{x}$  \citep{HsuBabin}. We find that $%
x\approx 0.7$, derived from observations , also provides a good fit to our
calculated profiles. However we also note that the departure between the
calculated and observed (fitted with the formula) profiles mainly comes from
the faster decay with $r$ of our profiles, at large $r$. This means that the
Eq.(\ref{eq}) generates maximal extension of the vortex that is somehow
shorter than that observed in reality. This is compatible with the
interpretation that the peripheral part of the tropical cyclone is dominated
by thermodynamic processes, which are absent from the FT description. For
the outer part of the tropical cyclone, \citet{Emanuel2004} considers local
balance between subsidence warming and radiative cooling. The radial
distribution of the azimuthal velocity results from the equality between the
Eckman suction and the subsidence rate. This strong thermodynamics aspect
goes beyond FT model (which relies on vorticity organization) and is the
main obstacle in verifying the FT result that $\left( L^{rad}\right)
^{phys}=R_{Rossby}$. We will then look for estimation of an "effective"
maximal radial extension (like the radius where the azimuthal velocity is $%
12\ \left( m/s\right) $) and we will evaluate the ability of the FT model to
describe the atmospheric vortex according to its ability to reproduce this
value.

\begin{figure}[tbph]
\centerline{\includegraphics[height=5cm]{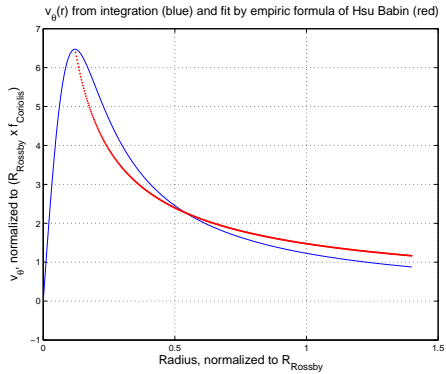}}
\caption{The profiles of the azimuthal velocity as results form integration of Eq.(3)
(blue) and from the empirical formula of Hsu - Babin (red) (\emph{i.e.} from
observations). The figure shows that the radial decay of the theoretical $%
v_{\theta }\left( r\right) $ at large $r$ gives systematically a shorter
maximum radial extension of the tropical cyclone.}
\label{fig11}
\end{figure}

\bigskip

\bigskip

\subsubsection{Calculation of the maximum radial extension of the tropical
cyclone using input from Shea - Gray and Chavas - Emanuel databases}

The database QuikSCAT of \citep{CE} (denoted CE in the following) is
organized as a collection of sets of several quantities measured for a
single observation on a tropical cyclone, in particular the maximum
velocity, the radius where the azimuthal velocity is $12\ \left( m/s\right) $%
, the maximum radial extension. The latter is obtained by extrapolation as
mentioned above. For almost all tropical cyclones in the CE database there
are sequences of observations at successive times, which we can use to
qualitatively identify stationarity plateaux, if any. Our results can only
be compared with such cases.

The data from SG and CE are used according to the following procedure.

(1) We read from CE, for a particular case (a line in the file), the maximum
velocity $\left( v_{\theta }^{\max }\right) _{CE}^{phys}\ \ \left(
m/s\right) $. (2) Using the fitting curve of SG we obtain $\left(
r_{v_{\theta }^{\max }}\right) _{SG}^{phys}\ \left( m\right) $. (3) Now we
turn to the two Eqs.(\ref{vmax} - \ref{rvmax}) and define an algebraic
equation whose solution is $R_{Rossby}$ corresponding to that particular
observation. (3.1) We start by assuming a value for $R_{Rossby}$ and with it
we normalize 
\begin{equation}
\left( v_{\theta }^{\max }\right) _{CE}^{phys}\ \ \left( m/s\right)
\rightarrow \left( v_{\theta }^{\max }\right) _{CE}=\frac{\left( v_{\theta
}^{\max }\right) _{CE}^{phys}}{R_{Rossby}f}  \label{normv}
\end{equation}%
\begin{equation}
\left( r_{v_{\theta }^{\max }}\right) _{SG}^{phys}\ \left( m\right)
\rightarrow \left( r_{v_{\theta }^{\max }}\right) _{SG}=\frac{\left(
r_{v_{\theta }^{\max }}\right) _{SG}^{phys}}{R_{Rossby}}  \label{normr}
\end{equation}%
(3.2) Next we ask that the velocity from CE, so normalized, equals the
velocity of Eq.(\ref{vmax})%
\begin{equation}
\left( v_{\theta }^{\max }\right) _{CE}=v_{\theta }^{\max }\left(
L^{sq}\right) =2.6461\times \exp \left( \frac{1}{L^{sq}}\right) -2.7748
\label{eq100}
\end{equation}%
This is an algebraic equation for $L^{sq}$. (4) The result is inserted in
Eq.(\ref{rvmax}) to determine $r_{v_{\theta }^{\max }}$, normalized. (5)
This must be compared with the \emph{normalized} value of the radius of
maximum velocity $\left( r_{v_{\theta }^{\max }}\right) _{SG}$ obtained from
SG, \emph{i.e.} Eq.(\ref{normr}). If they are different then we will change $%
R_{Rossby}$ and iterate (\emph{i.e.} return to 3.1) the sequence until the
equality is obtained. Therefore the equation to be solved is%
\[
\left. r_{v_{\theta }^{\max }}\left( L^{sq}\right) \right\vert _{Eq(\text{%
\ref{rvmax}})}=\frac{\left( r_{v_{\theta }^{\max }}\right) _{SG}^{phys}}{%
R_{Rossby}}
\]%
for the unknown $R_{Rossby}$. (6) Assuming that a solution exists, we find $%
\left( L^{sq}\right) ^{sol}$ (non-dimensional) and $\left( R_{Rossby}\right)
^{sol}$ . (7) Knowledge of these solutions allows to convert Eqs.(\ref{vmax}%
) and (\ref{rvmax}) into dimensional (physical) quantities that can be
compared with observations, other than those that have been involved in the
procedure described above. In particular, $r_{12}$, the radius where the
azimuthal velocity is $12\ \left( m/s\right) $, (from the database Chavas
Emanuel). We have chosen in CE a set of cases that seem to present
stationarity and carried out calculations. We illustrate the procedure in
the following three cases, with only the intention to clarify the procedure
explained above.

\paragraph{Case 1}

The position in CE database is Line \emph{440 BERTHA. }The latitude is $%
\theta =29.65$ and the Coriolis parameter is%
\[
f=2\Omega \sin \theta =7.1951\times 10^{-5}\ \left( s\right) 
\]%
From CE we take $\left( v_{\theta }^{\max }\right) _{CE}^{phys}=40.098\
\left( m/s\right) $. It results%
\begin{eqnarray*}
R_{Rossby} &=&178862\ \left( m\right) \\
L^{sq} &=&1.2427
\end{eqnarray*}

Now we can make further comparison with observations, in particular with the
radius of, $v_{12}$, \emph{i.e.} $v_{\theta }=12\ \left( m/s\right) $, which
in CE is $\left( r_{v_{\theta }=12}\right) _{CE}^{phys}=166411\ \left(
m\right) $Since now we know $R_{Rossby}$ we normalize the velocity with $%
R_{Rossby}\times f=12.86\ \left( m/s\right) $,%
\[
\frac{v_{12}}{12.86}=\frac{12}{12.86}=0.9331 
\]%
We return to solve Eq.(\ref{eq}) for $L^{rad}=L^{sq}\sqrt{2}=1.7574$ and
find the radial profile of the (normalized) velocity, $v_{\theta }\left(
r\right) $. On this profile, $v_{\theta }=0.9331\ $ is found at $%
r_{v=0.9331}\approx 1.26$ which means%
\begin{equation}
\left( r_{v=0.9331}\right) ^{phys}=178862\times 1.26\ \left( m\right)
=225370\ \left( m\right)  \label{eq102}
\end{equation}%
This compares well with the data from CE $\left( r_{v_{\theta }=12}\right)
_{CE}^{phys}=225129\ \left( m\right) $.

For the maximum radial extension we find 
\begin{equation}
R_{\max }=L^{sq}\sqrt{2}\times R_{Rossby}=314340\ \left( m\right)
\label{eq104}
\end{equation}%
which is again small compared with the data from CE $R_{\max }^{CE}=391874\
\left( m\right) $.

Note similarity with \emph{625} \emph{MARTY}.

\paragraph{Case 2}

This is \emph{480 OMAR}. The latitude is $\theta =16.44$ and the Coriolis
frequency is $f=2\Omega \sin \theta =4.1162\times 10^{-5}\ \left(
s^{-1}\right)$

From CE we take $\left( v_{\theta }^{\max }\right) _{CE}^{phys}=46.27\
\left( m/s\right) $. It results following the procedure described above%
\begin{eqnarray*}
R_{Rossby} &=&202193\ \left( m\right) \\
L^{sq} &=&0.87
\end{eqnarray*}

Now we can make further comparison with observations, in particular with the
radius of, $v_{12}$, \emph{i.e.} $v_{\theta }=12\ \left( m/s\right) $, which
in CE is $\left( r_{v_{\theta }=12}\right) _{CE}^{phys}=187613\ \left(
m\right) $ Since now we know $R_{Rossby}$ we normalize the velocity with $%
R_{Rossby}\times f=8.32\ \left( m/s\right) $,%
\[
\frac{v_{12}}{8.32}=\frac{12}{8.32}=1.4418 
\]%
We return to solve Eq.(\ref{eq}) for $L^{rad}=L^{sq}\sqrt{2}=1.2304$ and
find the radial profile of the (normalized) velocity, $v_{\theta }\left(
r\right) $. On this profile, $v_{\theta }=1.4418\ $ is found at $%
r_{v=1.4418}\approx 0.9$ which means%
\begin{equation}
\left( r_{v=0.9331}\right) ^{phys}=202193\times 0.9\ \left( m\right)
=181970\ \left( m\right)  \label{eq106}
\end{equation}%
This compares well with the data from CE $\left( r_{v_{\theta }=12}\right)
_{CE}^{phys}=187613\ \left( m\right) $

For the maximum radial extension we find $R_{\max }=L^{sq}\sqrt{2}\times
R_{Rossby}=248770\ \left( m\right)$ which is again small compared with the
data from CE $R_{\max }^{CE}=423035\ \left( m\right) $.

We note however that for similar data (line 509 Aletta of CE) with $\left(
v_{\theta }^{\max }\right) _{CE}^{phys}=46.205\ \left( m/s\right) $ at
latitude $14.68$, we have $f=3.7129\times 10^{-5}\ \left( s^{-1}\right) $
and obtain $R_{Rossby}=213070\ \left( m\right) $ and $L^{sq}=0.8458$. After
similar calculations we get $\left( r_{v_{\theta }=12}\right)
^{phys}=191763\ \left( m\right) $ while $\left( r_{v_{\theta }=12}\right)
_{CE}^{phys}=122620\ \left( m\right) $. The difference is substantial. While
the calculation, for close magnitudes, gives close results, the reality (%
\emph{i.e.} the observation) may be rather different: close magnitudes of $%
v_{\theta }^{\max }$ and of $f\left( \theta \right) $ can give very
different $r_{12}$'s.

\paragraph{Case 3}

This is the line \emph{299 KARL} in the CE database.\emph{\ }The input is $%
\left( v_{\theta }^{\max }\right) _{CE}^{phys}=48.92\ \left( m/s\right) $
Since the \emph{latitude is }$15^{o}$, we have a Coriolis frequency (taking $%
\Omega =7.2722\times 10^{-5}\ \left( s^{-1}\right) $) 
\[
f=2\Omega \sin \theta =3.7644\times 10^{-5}\ \left( s^{-1}\right) 
\]

Using $\left( v_{\theta }^{\max }\right) _{CE}^{phys}$ we start a search of $%
R_{Rossby}$. For every step, using the current guess for $\left(
R_{Rossby}\right) ^{\left( k\right) }$ we convert to non-dimensional velocity%
\[
\frac{\left( v_{\theta }^{\max }\right) ^{CE}}{R_{Rossby}^{\left( k\right)
}\times f} 
\]%
and impose to be equal to Eq.(\ref{vmax}), which determines $\left(
L^{sq}\right) ^{\left( k\right) }$. With\emph{\ }$\left( v_{\theta }^{\max
}\right) _{CE}^{phys}$ we calculate by spline interpolation on the \textbf{%
Shea Gray} data, $\left( r_{v_{\theta }^{\max }}\right) _{SG}\ \ \left(
m\right) $ and normalize 
\[
\frac{\left( r_{v_{\theta }^{\max }}\right) _{SG}}{R_{Rossby}^{\left(
k\right) }} 
\]%
This is compared with $r_{v_{\theta }^{\max }}$ from Eq.(\ref{rvmax}) where $%
\left( L^{sq}\right) ^{\left( k\right) }$ has been inserted. The comparison
is used as equation and an iterative procedure (the NAG subroutine \textbf{%
c05awf} is employed) leads to the solution. It resulted%
\begin{equation}
R_{Rossby}=196554\ \left( m\right) ,L^{sq}=0.7891  \label{eq108}
\end{equation}

This corresponds to $L^{rad}=1.1126$. We now want to estimate the radius
where the azimuthal velocity takes value $12\ \left( m/s\right) $, using the
approach based on FT. We first normalize the velocity, since $R_{Rossby}$ is
known 
\begin{eqnarray*}
R_{Rossby}\times f &=&196554\times 3.7644\times 10^{-5}=7.39\ \left(
m/s\right)  \\
v_{12} &=&\frac{12\ \left( m/s\right) }{R_{Rossby}\times f}=1.6218
\end{eqnarray*}%
We find the radial profile of the azimuthal velocity by performing the
radial integration of Eq.(\ref{eq}) with $L^{rad}=L^{sq}\sqrt{2}=1.1126$. On
the plot $\left( r,v\right) $ the velocity $v_{12}=1.6218$ is obtained at
the radius $r_{v=1.6218}\approx 0.85$. Now we can return to dimensional
quantities%
\begin{equation}
\left( r_{v=1.6218}\right) ^{phys}=R_{Rossby}\times 0.85=167070\ \left(
m\right)   \label{eq110}
\end{equation}%
This is smaller than the value found in CE, $\left( r_{12}\right)
_{CE}=206639\ \left( m\right) $, a possible reflection of the weak ability
of FT to describe the peripheral region of the vortex, where thermodynamics
is stronger. The estimation for the maximum radial extension is%
\[
R_{\max }=L^{sq}\sqrt{2}\times R_{Rossby}=218690\ \left( m\right) 
\]

\subsection{General conclusion of numerical verifications}

The numerical study of the implications of the FT model is not the subject
of this work and we only refer to results that involve the equality $%
R_{Rossby}=R_{\max }$. After a large number of numerical applications of Eq.(%
\ref{eq}) and of its consequences, Eqs.(\ref{vmax}) and (\ref{rvmax}),
including those of \citet{FlorinMadiGAFD}, we note a general trend. The FT
model of the vorticity self-organization leads to vortical structures that
have high maximal azimuthal wind for small spatial extension of the
atmospheric vortex. For this reason there are cases where the results of the
model are sensibly different from observations, \emph{i.e.} the calculated
maximal radial extension is smaller than that observed, $R_{\max
}^{calc}<R_{\max }^{obs}$. Since the calculated vorticity is almost zero
towards the peripheral region, one may expect that the thermal processes,
which we cannot include, are dominant in that region. 

The calculations also associate high azimuthal wind with small eye-wall
radius and this is compatible the observations, as shown for example by the
empirical formula of Willoughby and Rahn \citeyear{WilloughbyRahn1}. 

Finally, we note the high sensitivity (already mentioned previously %
\citep{FlorinMadiGAFD}) of the results to even small variation of the input
data. This is clearly seen in the two equation (\ref{vmax}) and (\ref{rvmax}%
) where the dependence on the parameter $L^{sq}$ is exponential.

There are many cases where Eq.(\ref{eq}) and Eqs.(\ref{vmax}), (\ref{rvmax})
are close to the observed values and in general they are never far from
reality. This may be the signature that the self-organization of the
vorticity has a substantial role in the stationary state of the tropical
cyclone.

\bigskip

\section{Conclusions}

In a purely theoretical framework and on very general basis we have derived
the equality between the maximum radius $R_{\max }$ of the tropical cyclone
and the Rossby radius $R_{Rossby}$. However this has been done by only
taking into account the process of vorticity self-organization. Since this
is just a part of the full dynamics of a tropical cyclone, we cannot expect $%
R_{\max }=R_{Rossby}$ to be an exact result. A priori we do not know how much
of the full dynamics is influenced by the strictly kinematic organization of
the elements of vorticity. Qualitatively, the predominance of the intrinsic
self-organization of vorticity (over the source/sink processes) may exist
close to the stationarity and within the validity of the two-dimensional
approximation. During cyclogenesis there is continuous generation of
vorticity and continuous process of organization. Close to stationarity the
rate of generation of new vorticity is reduced but the process of
organization continues. We note that a detailed feature (not discussed here)
of the FT formulation reveals that there is no exact stationarity and the
vorticity concentration actually continues at very small rate.

This property of the large scale stationary atmospheric vortex, $%
R_{max}=R_{Rossby}$, has been derived from the mapping that connects the
vorticity self-organization to the extremum of an action functional
(integral of the Lagrangian density Eq.(\ref{26})), a classical field
theory. The equality of the masses of the matter field particle and of the
gauge particle translates through the mapping into the equality of the
radial extension of the vortex with the Rossby radius. In numerical
calculations this relationship is used either directly or implicitly, 
\textit{i.e.} comparing with observation some important characteristics of
the tropical cyclone (maximum velocity, radius of the maximum velocity,
maximum radial extension). In cases that can be used within our
approximations, the FT model (implicitly $R_{max}=R_{Rossby}$ ) reproduces
reasonably well the observation.

Comparing our results with observation takes then a particular meaning: we
actually obtain an idea about the importance of the vorticity
self-organization within the full dynamics. It suggests that the process of
self-organization of the vorticity, a part of the dynamics of the tropical
cyclone that is distinct of any thermodynamic process, appears as an
important factor determining the spatial distribution of the main flow
variables. It seems to become a necessity to combine the spontaneous
self-organization with the thermodynamics of the atmospheric vortex. This is
an important area of investigation.

\bigskip

\textbf{Aknowledgments}. This work has been partly supported by the Grant
ERC - Like 4/2012 of UEFISCDI. The authors aknowledge useful discussions
with Jun - Ichi Yano, Robert S. Plant and Emmanuel Vincent. The fruitful exchange of
ideas within the COST collaboration ES0905 is highly appreciated.

\bigskip

\bibliography{worknodoi}

\end{document}